\documentclass[12pt]{article}

\usepackage{scicite}

\usepackage{times}
\usepackage{graphicx}

\newenvironment{sciabstract}{%
\begin{quote} \bf}
{\end{quote}}

\title{Gamma-ray Bursts as distance indicators through a machine learning approach}

\author
{Maria Dainotti$^{1,2,3,4}$, Vah\'e Petrosian$^{1,5}$, Malgorzata Bogdan$^{6,7}$, \\ Blazej Miasojedow$^{8}$, Shigehiro Nagataki$^{9,10}$, Trevor Hastie$^{11}$, \\Zooey Nuyngen$^{12}$, Sankalp Gilda$^{13}$, Xavier Hernandez $^{14}$, Dominika Krol$^{2}$\\
\\
\normalsize{$^{1}$Department of Physics and KIPAC, Stanford University,Stanford, CA 94305, USA}\\
\normalsize{$^{2}$Jagiellonian University, Orla 171, Krakow, Poland}\\
\normalsize{$^{3}$Space Science Institute, 4750 Walnut St, Suite 205, Boulder, CO, 80301, US}\\
\normalsize{$^{4}$INAF, Via Pietro Gobetti, Bologna, Italy}\\
\normalsize{$^{5}$Department of Applied Physics, Stanford University, Stanford, CA 94305, USA}\\
\normalsize{$^{6}$Faculty of Mathematics and Computer Science, University of Wroclaw, Poland}\\
\normalsize{$^{7}$Department of Statistics, Lund University, Sweden}\\
\normalsize{$^{8}$Faculty of Mathematics, Informatics and Mechanics, University of Warsaw, Poland}\\
\normalsize{$^{9}$Astrophysical Big Bang Laboratory, RIKEN, Saitama 351-0198, Japan}\\
\normalsize{$^{10}$Interdisciplinary Theoretical \& Mathematical Science Program (iTHEMS),} \\ \normalsize{ RIKEN, Saitama
	351-0198, Japan}\\
\normalsize{$^{11}$Department of Statistics, Stanford University, Stanford, CA 94305, USA}\\
\normalsize{$^{12}$Department of Astronomy, University of Florida, Gainesville, FL, 32611, US}\\
\normalsize{$^{13}$Instituto de Astronomia, Universidad Nacional Autonoma de Mexico,}\\ \normalsize{Apartado Postal 70–264, C.P. 04510, Mexico D.F., Mexico}\\
\normalsize{$^{14}$Department of Astronomy, University of California Los Angeles, UCLA, Los Angeles, CA, US}\\
\\
\normalsize{$^\ast$E-mail:  mdainott@stanford.edu.}
}

\date{}

\begin{document} 


\baselineskip24pt


\maketitle 


\begin{sciabstract}
Gamma-ray bursts (GRBs) are spectacularly energetic events, with the potential to inform on the early universe and its evolution, once their redshifts are known. Unfortunately, determining redshifts is a painstaking procedure requiring detailed follow-up multi-wavelength observations often involving various astronomical facilities, which have to be rapidly pointed at these serendipitous events. Here we use Machine Learning algorithms to infer redshifts from a collection of observed temporal and spectral features of GRBs. We obtained a very high correlation coefficient ($0.96$) between the inferred and the observed redshifts, and a small dispersion (with a mean square error of $0.003$) in the test set. The addition of plateau afterglow parameters improves the predictions by $61.4\%$ compared to previous results. The GRB luminosity function and cumulative density rate evolutions, obtained from predicted and observed redshift are in excellent agreement indicating that GRBs are effective distance indicators and a reliable step for the cosmic distance ladder.
\end{sciabstract}


\section*{Introduction}\label{intro}

The localization of some GRBs in host galaxies established directly their extragalactic origin, and by now redshifts, $z$, have been measured for several hundred GRBs, with a several having $z$ between $6$ and $9.4$. As a result, there have been numerous efforts to use GRBs as cosmological probes for shedding light on astrophysical processes at these early stages of the evolution of the universe, to use them to determine the evolution of their luminosity function, to assess if the GRB rate follows the star formation rate (SFR) at low redshifts, and in some cases using them as ``standard candles", through correlations between distance dependent and distance independent variables, for cosmological studies. A persistent caveat on the use of GRB correlations for cosmological studies is the incompleteness of the samples of GRBs with known redshifts. In addition, most observed and derived characteristics of GRBs such as peak prompt luminosity, $L_p$, duration, e.g.~$T_{90}$, gamma-ray spectral parameters (power-law indices and peak photon energy), and several afterglow features have broad intrinsic distributions. This has made it extremely difficult to find an effective strategy to reliably derive redshifts from the more readily available observational features. In addition, it is essential to determine the variations with redshift of these distributions, namely their cosmological evolution, before attacking any of the above problems.
This task, however, requires large samples of GRBs with known redshifts, well defined observational selection criteria, and robust data analysis that will allow a more precise determination of the cosmological evolutions at play. 

Various approaches have been used to determine the evolution of the luminosity function and the density rate evolution. These fall into two broad categories: parametric forward fitting and non-parametric. In the former, a set of assumed {\it parametric} functional forms are fit to the  data to determine the ``best fit values" of the relevant parameters. This involves assumed forms for several functions, such as the luminosity function, luminosity and formation rate evolutions, photon spectrum, light curve, etc., each with three or more parameters. Because of the large number of parameters, there is no certainty on the uniqueness of the results. Moreover, these methods often require binning of the data and hence large samples. There have been many such studies of both long GRBs (see e.g., Porciani \& Madau 2001; Jakobson et al. 2006; Salvaterra et al. 2009, Madau, P. \& Dickinson, M. 2014, and references therein) and short GRBs (Guetta \& Piran 2005, 2006; Nakar \& Piran 2005; Metzger \& Berger 2012; Petrillo et al. 2013; Ghirlanda et al. 2016, Wang et al. 2017). 

There are various {\it non-parametric, non-binning methods} that are more powerful, especially for small samples. 
For example, Schmidt (1968) using the so-called $V/V_{\rm max}$ method on a sample of 33 quasi-stellar radio sources, with one-sided truncation (due to a lower limit on the flux) was able to determine their luminosity function and density evolution.
A review by Petrosian (1992) describes further development of this and other similar methods with
the conclusion that all these methods are equivalent to the more general Lynden-Bell (1971) $C^{-}$
method. However, as also pointed out in this review, all these methods require the
critical assumption that the variables are uncorrelated, or independent.
This shortcoming led to the development of the more powerful (also non-parametric,
non-binning) methods by Efron \& Petrosian (1992; EP), which first
tests for the correlation between the variables. If correlations are found, it
introduces a variable transformation that yields an uncorrelated set of variables. It then
proceeds with the determination of the distribution of the uncorrelated variables using the $C^{-}$ or other
non-parametric method \footnote{This approach was later generalized to two-sided truncated data (e.g.~data
with upper and lower flux limits) in Efron \& Petrosian (1999)}. This combined 
Efron-Petrosian and Lynden-Bell method (EPL) has been
proven to be  very useful for studies of many aspects of Active Galactic Nuclei
(see, e.g. Maloney \& Petrosian 1999; Singal et al. 2011) and long GRBs (Lloyd \&
Petrosian, 1999; Lloyd et al. 2000 and 2002; EP04;  Kocevski \& Liang 2006; Yonetoku, et
al. 2004; 2014; Dainotti et al. 2013a, 2015b, 2017a; Petrosian et al. 2015; Yu et al. 2015; Pescalli et
al. 2016; Tsvetkova et al. 2017). A very interesting aspect of the last four papers has been 
the finding of a distinct difference between the density rate evolution of the long GRB and global star formation rates at low redshifts ($0<z<1$). The differences in these results could well be due to the fact that the samples used in these papers are complete to different percentages and to different flux limits. For example, Petrosian et al. (2015) used a more conservative gamma-ray threshold to assure a better completeness of the sample. Pescalli et al. (2016) use an even higher flux limit $f_{lim}=5 \times 10^{-8} \rm erg$ $\rm cm^{-2} s^{-1}$ than the ones used in the other $3$ papers, and obtain a smaller difference between the GRB and SFRs, but their sample is only $85\%$ complete. 

Our goal in this paper is to increase the number of GRBs with inferred redshift considerably. This method will allow to more than double or even triple the sample of {\it Swift (Gehrels et al. 2004) GRBs} with $z$, so that we can finally solve the above and other ongoing debates on the nature of GRBs.
The situation is the same also for short GRBs where there has only been some preliminary investigation of their luminosity function and cosmological evolution. An accurate determination of these for short GRBs is becoming more crucial with the dawn of gravitational wave (GW) astronomy.
Therefore, having a method that can reliably estimate redshifts for more GRBs is of paramount important. The direct determination of the redshift of a GRB requires rapid localization and spectral observations. {\it Swift}, like its predecessor {\it Beppo-SAX}, uses X-ray instruments for localization, but unlike {\it Beppo-SAX}, which relied on ground based observations of host galaxies, {\it Swift} can
obtain spectra and sometimes redshifts with the on-board UVOT instrument. As a result {\it Swift} has been
able to secure redshifts for a larger fraction of the GRBs it detects. Still, this fraction remains small ($\sim 30\%$).

As mentioned above, using correlations (Amati et al. 2002, Ghirlanda et al. 2004, Yonetoku et al. 2004, Dainotti et al. 2008) between a distance independent characteristic (e.g. time at the end of the plateau of the afterglow emission, the peak photon energy in the $\nu F_\nu$ spectrum, etc.) and a distance dependent one (the luminosity at the end of the plateau emission, isotropic prompt peak luminosity and energy) has resulted in attempts to obtain {\bf pseudo-redshifts} for larger and possibly more complete samples for cosmological studies (Lloyd-Ronning \& Ramirez-Ruiz 2002; Atteia et al. 2003; Yonetoku et al. 2004; Kocevski \& Liang 2006; Dainotti et al. 2011a, Zhang \& Wang 2018). However, using linear relations between relevant GRB parameters to infer redshifts has not yet led to accurate estimates because all these relations depend on the luminosity distance ($D_L$) which itself depends on $z$ and cosmological parameters in a complex way. In addition, a small error in the determination of $D_L(z)$ implies a large uncertainty in the determination of $z$ at high redshifts. So, the reliability of such redshifts is questionable. Instead, here we propose a method that bypasses completely the determination of the luminosity distance. More specifically, we explore the use of {\it machine learning} (ML) procedures for the determination of redshifts, with training and validation performed for a sample of 171 GRBs with known redshift. 

In the next section (\S The sample) we describe the GRB sample which we use to develop a redshift inference procedure using ML methods described in \S Methods, where we describe the SuperLearner method that we have applied to the chosen sample, comprising several ML algorithms. In \S results we describe how well the chosen procedures reproduce the observed spectroscopic redshifts of the sample, where we find an excellent correlation between the predicted and measured redshifts, with a very small dispersion. In \S luminosity function and cumulative rate evolution we compare the results from application of the EPL methods to both sets of redshifts (predicted and observed) to obtain the luminosity evolution, luminosity function and cumulative density rate evolution. An excellent agreement between both of the above samples further validates the method.
A brief summary and discussion of the impact on future work are given in \S summary and conclusion.

\section*{The sample}\label{sample}

\begin{figure}
\includegraphics[width=1.0\linewidth,height=0.80\textwidth,angle=0,clip]{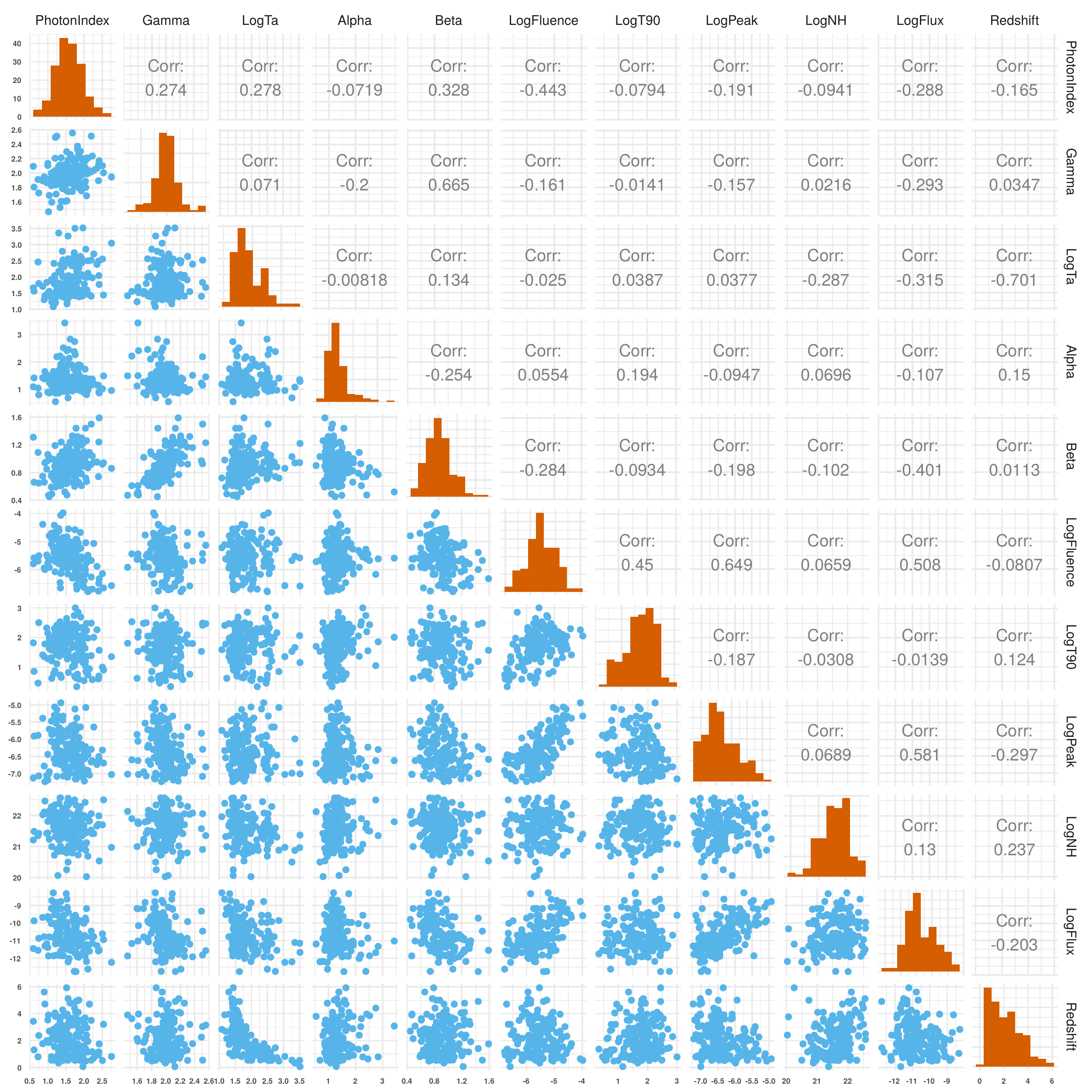}
\caption{Scatter matrix plot of the all variables at play, some showing strong correlations, while others show weak or no correlation. The scatter matrix is divided in two parts, the bottom part shows the logarithms of the prompt Fluence, $T_{90}$, the prompt peak flux, the plateau X-ray flux, $F_{X,a}$, its duration, $T_{a}$, the hydrogen column density, $NH$, and the linear distributions of the prompt photon index, the photon and spectral indices, $\Gamma$, and $\beta$, of the X-ray afterglow and plateau emissions, respectively, and the temporal decay index after the plateau, $\alpha$. Each variable is plotted against every other. On the diagonal, the differential distribution is shown for each variable, and numbers in the upper part of the panel provides the Pearson correlation coefficient, $r$, between variables.}
\label{scatter}
\end{figure}

Our analysis adopts sets of several observed characteristics of GRBs, which we treat as {\it explanatory
variables} or {\it predictors}, from both the prompt and the afterglow (in particular the plateau)
emission, to be used in the prediction of their redshifts. Both data sets are from the {\it Swift} GRB catalog, more precisely the BAT+XRT observations, covering the period from January $2005$ to January $2019$. 
Our general aim is to develop a methodology to reliably infer redshifts from such observations. In this paper we use a sample of GRBs with reliable known redshifts (redshifts which are not precisely determined are not included) and demonstrate how well the predicted redshifts agree with the observed ones.  
Fig. \ref{scatter} shows a scatter diagram of all the used variables (including the redshift) against each other. We have removed $5\%$ outliers (see below) since they may not be representative of the population and could affect the prediction of the redshift. This matrix includes the logarithms of the observed redshift, the prompt Fluence, $T_{90}$, the prompt peak flux, the plateau X-ray flux, $F_{X,a}$, its 
duration, $T_{a}$, and the hydrogen column density, $NH$, and the following variables in linear scale: the prompt photon index, the photon and spectral indicies (assuming a simple power-law spectrum), $\Gamma$, and $\beta$, of the X-ray afterglow and plateau emissions, respectively, and the temporal decay power law index after the plateau, $\alpha$. 

As it is clear from the Fig. \ref{scatter}, some of these scatter diagrams show clear correlations with each other, which are related
to phenomenological relations, such as the tri-dimensional Dainotti et al. (2016, 2017c) relation between the plateau luminosity, the rest frame duration, $L_a$ and $T^{*}_a$, and the prompt peak luminosity, $L_{peak}$ and the various prompt spectral hardness- luminosity relations (Collazzi \& Schaefer 2008); for recent reviews see Dainotti \& Del Vecchio (2017b); Dainotti et al (2018); and Dainotti \& Amati (2018). In addition to these relations, we also note a strong correlation between prompt fluence and peak flux, and between afterglow indecies $\Gamma$ and $\beta$.
The selection of the variables and the selection cuts is based on our knowledge of the parameters at play. For example, since we are using data from {\it Swift} BAT ($15-150$ keV) energy range we do not use the peak energy, $E_{peak}$, values since for the majority of cases the determination of $E_{peak}$ may not be reliable due to the small energy range sampled. We also performed the following selections cuts: $T_a \geq 10$ seconds, since values smaller than this limit are fitting error dominated so that no plateau feature can be reliably identified. Non-physical values of $\beta < 0$ and $\Gamma < 0$, and $\log NH < 20$ were also removed. Since long and short GRBs have different progenitors, as confirmed by the discovery of the gravitational wave associated with GRB 170817 (Abbott et al. 2017, Troja et al. 2017), we choose to perform the analysis using only long GRBs (with $T_{90} \ge 2$ s, see Mazets \& Golenetskii 1988, Kouveliotou et al. 1993).
To additionally remove outliers that might not be representative of the full sample, we applied the procedure of principal component analysis (PCA) and fitted a multivariate normal distribution to the set of principal components. We eliminated GRBs which fall in the $5\%$ tail of this distribution, so that we are confident that our sample is statistically representative of the full distributions of all parameters. 

\section*{Results}\label{results}
We started with a sample of $229$ GRBs, detected from January 2005 to January 2019, after the first selection cuts described above we retain $185$ GRBs, and after the application of the PCA cuts we are left with a sample of $171$ GRBs shown in Fig. \ref{scatter}.  

We show the results derived with the ensamble that uses the ML methods discussed above.
Results show that a model produced by the generalized additive model (GAM) has the highest predictive power (coefficient of $A_1=0.61$) followed by Extreme Gradient Boosting ($A_2=0.29$) and LASSO for linear models with interaction ($A_3=0.10$) where $A_1$, $A_2$ and $A_3$ are the relative merits of the models. The condition of the normalization of the coefficients is that $\sum {A_i}=1$. The metric we use to define the goodness of our results is based on the minimization of the root mean square error in the SuperLearner algorithm. We additionally use: the Pearson correlation coefficient, $r$, among $z_{obs}$ and $z_{pred}$, the mean squared error, MSE, defined as the $<(\log (z_{i,predicted}+1) - \log (z_{i,observed}+1)^2>$ where with the symbol $<>$ we indicate the average, and the bias as $<(\log (z_{i,predicted}+1)- \log z_{i,observed}+1)>$.
 
\begin{figure}
\includegraphics[width=0.50\linewidth,height=0.50\textwidth,angle=0,clip]{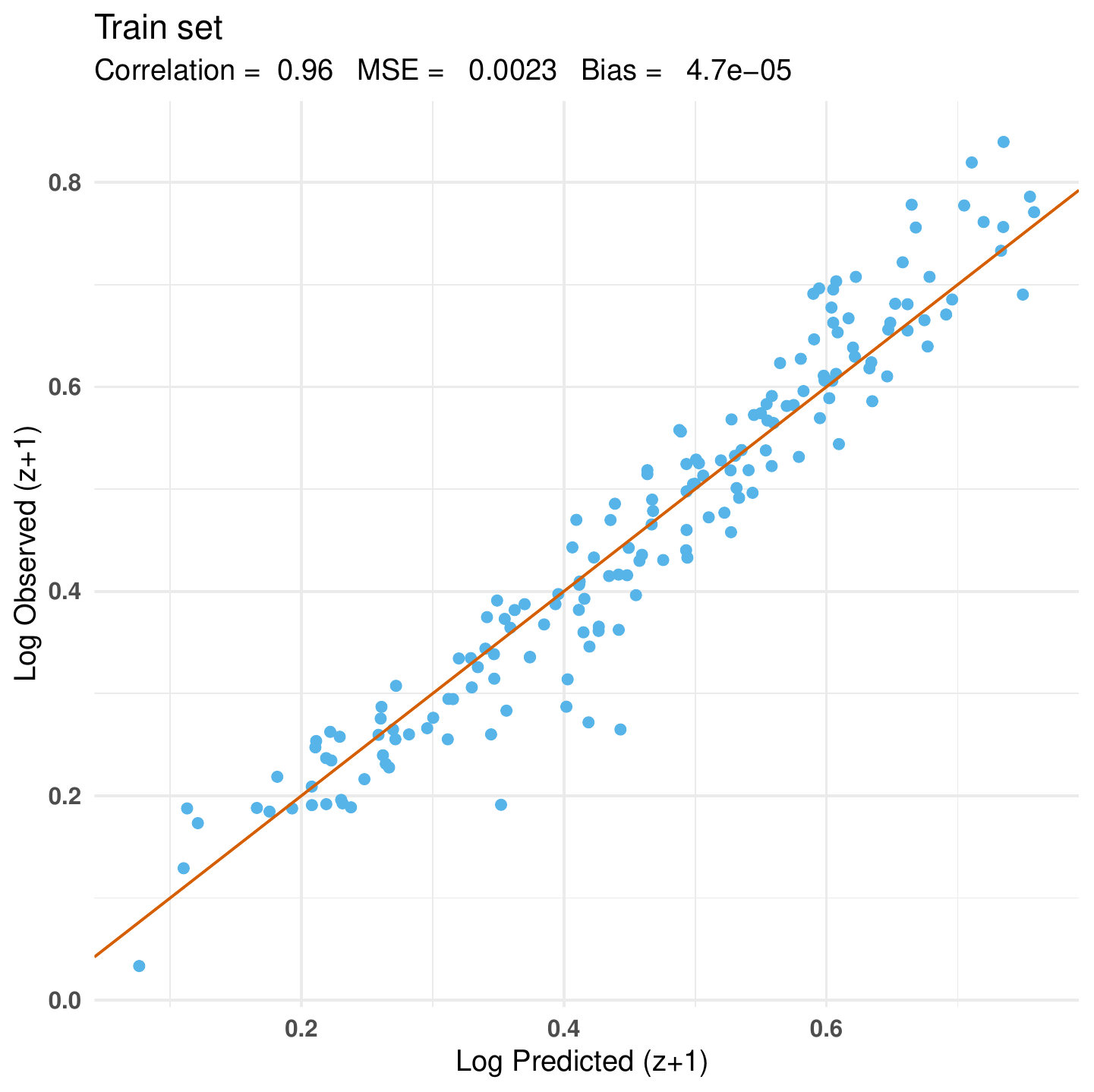}
\includegraphics[width=0.50\linewidth,height=0.50\textwidth,angle=0,clip]{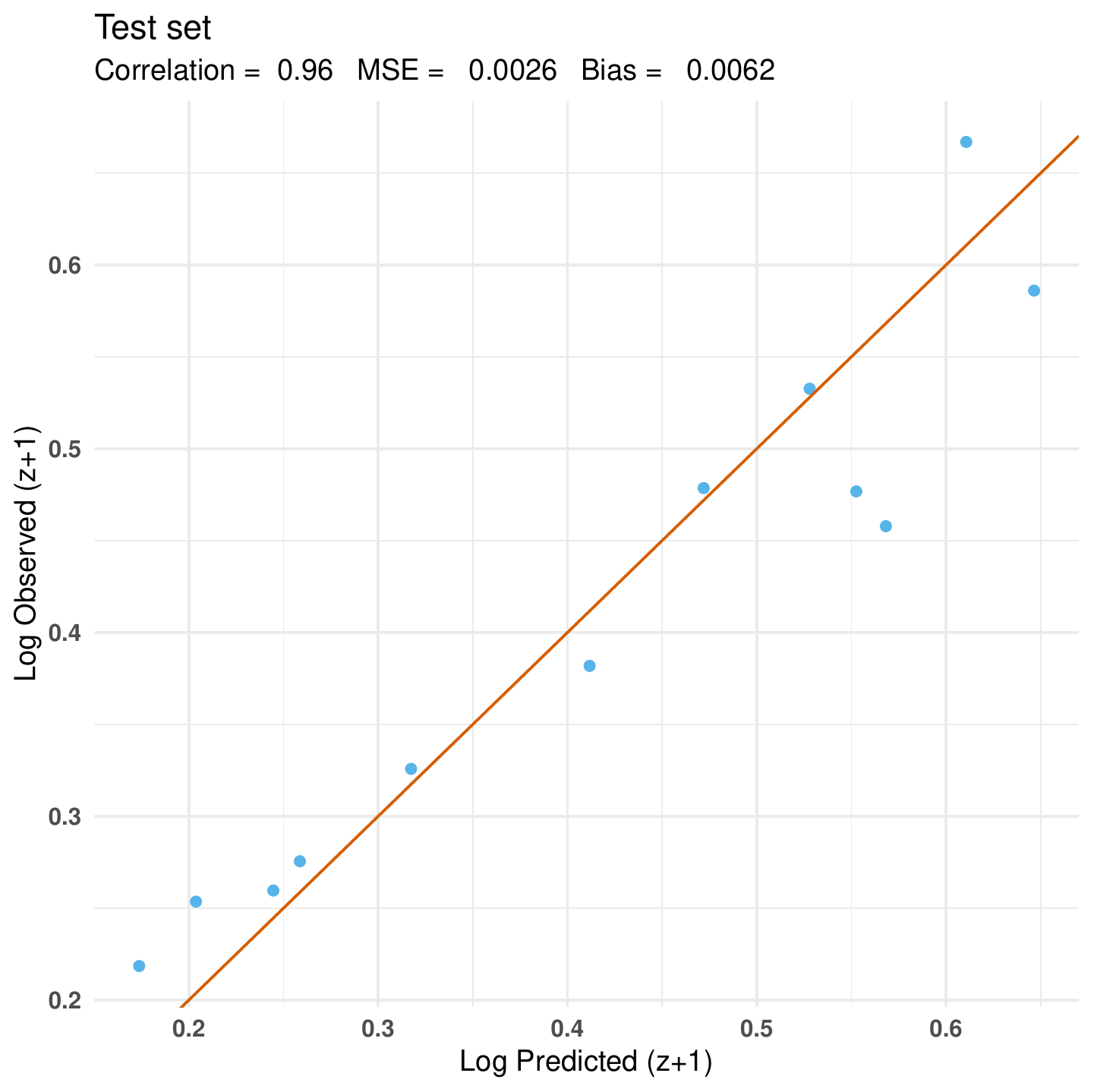}
\caption{Left panel: Predicted vs observed redshifts using SuperLearner on the training set of long 159 GRBs collected from January 2005 until January 2017. Right panel: Predicted vs observed redshifts using SuperLearner on the test set data (12 long GRBs collected from February 2017 until January 2019).}
\label{ensambletraining}
\end{figure}

The left panel of Fig. \ref{ensambletraining} shows the redshift predicted when a {\it training set} consisting of $90\%$ of the total sample of 171 GRBs collected from January 2005 until January 2017, namely 159 GRBs. The right panel shows how the prediction works on the {\it test set} (GRBs collected from February 2017 until January 2019, namely 12 GRBs), remaining $10\%$ of the total sample. We can see that for the training set the results are an excellent reproduction of the observed redshifts, as also happens with the test set. 
More specifically, for the training set we obtained $MSE=2.2 \times 10^{-3}$, $r=0.96$ and the bias$=4.7 \times 10^{-5}$, for the test set $MSE=3\times 10^{-3}$, $r=0.96$ and the bias $=6 \times 10^{-3}$, respectively. It is clear that we consistently obtain high values of $r$ and low values of $MSE$ and the bias for both samples indicating that we are not overfitting. 

\begin{figure}
\includegraphics[width=8.3cm,height=7.2cm,angle=0,clip]{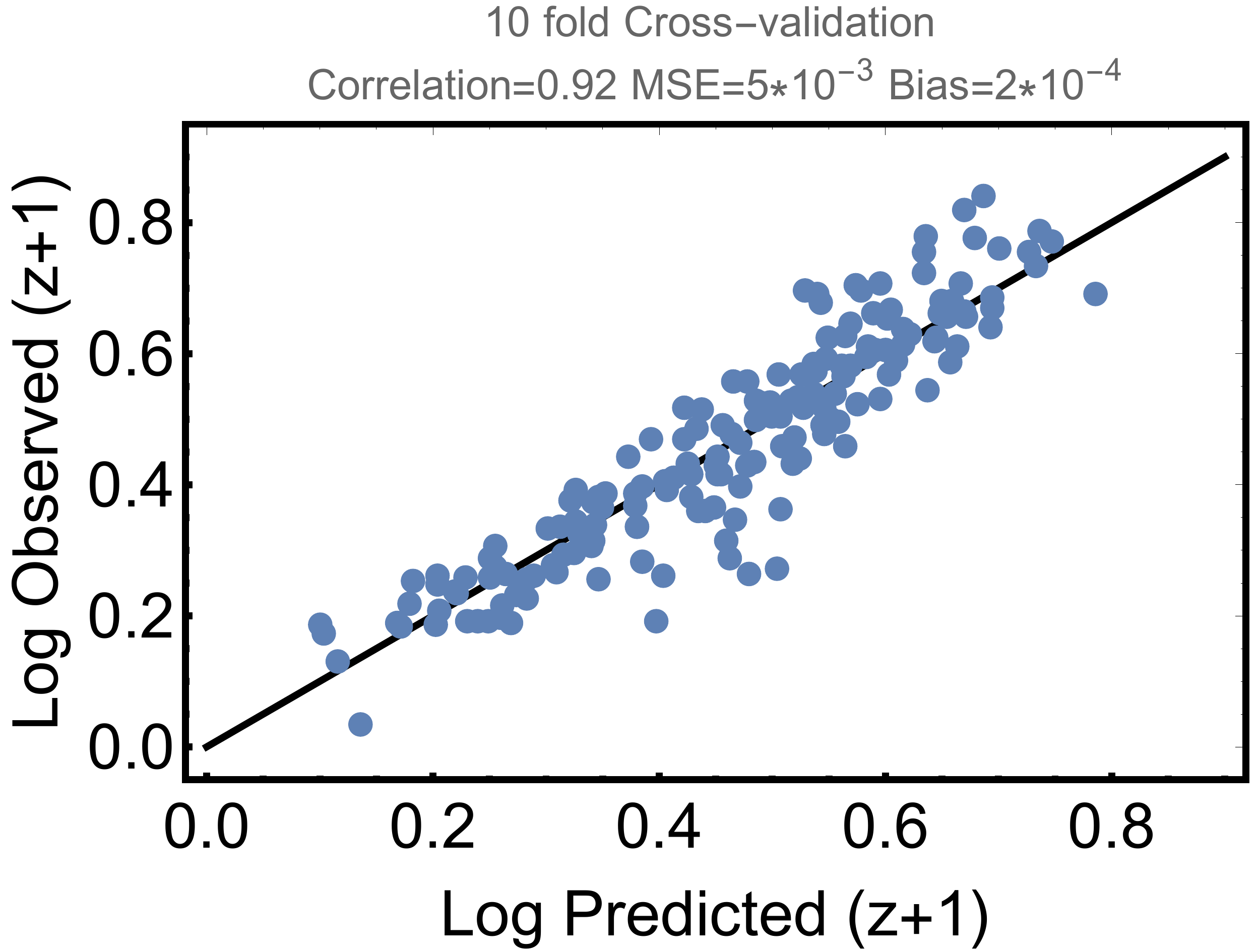}
\includegraphics[width=8.3cm,height=7.2cm,angle=0,clip]{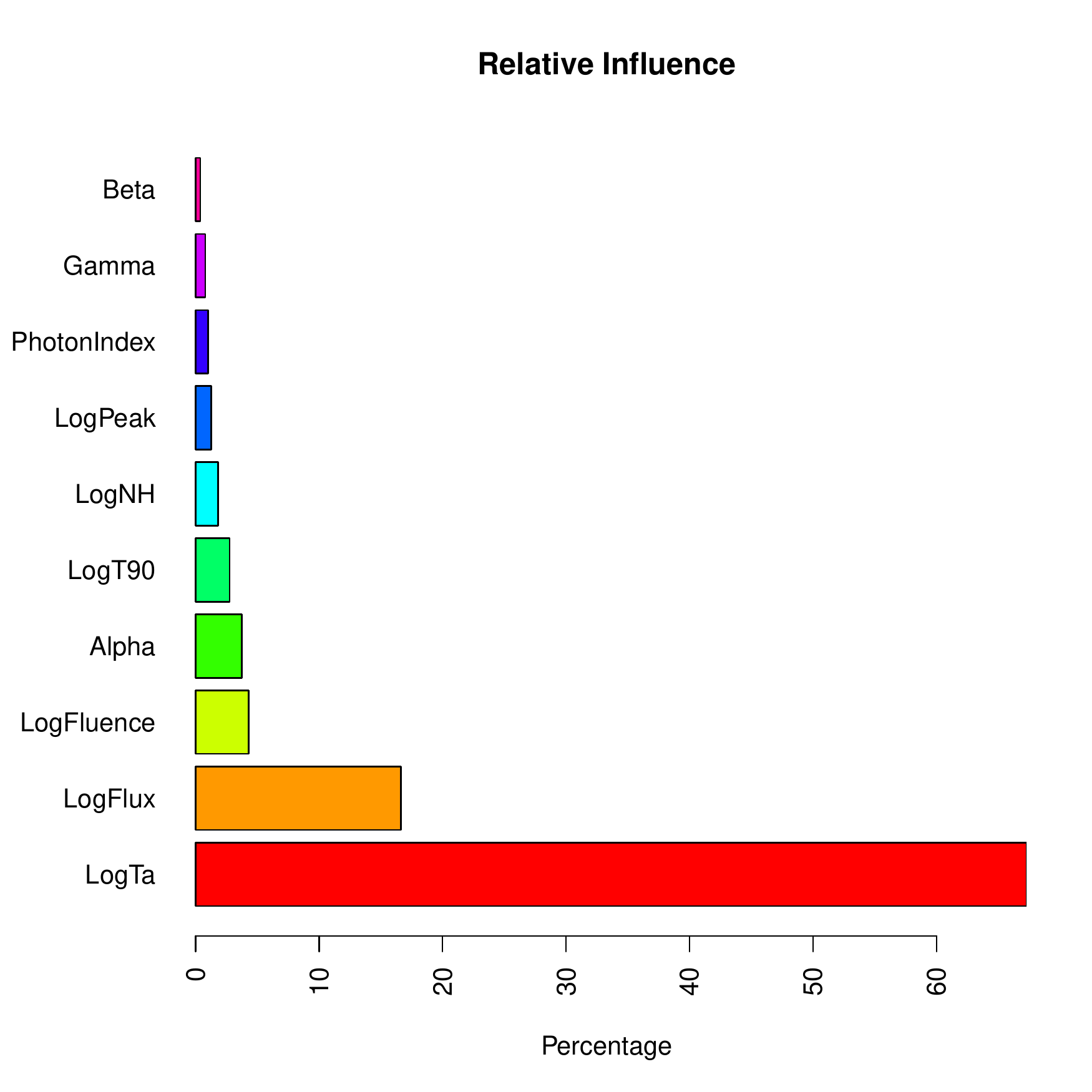}
\caption{Left panel: Predicted vs observed redshifts using SuperLearner on the full set of 171 GRBs.
Right panel: The relative influence of each variable on the predicted results. As expected the flux and duration of the plateau of the afterglow and the prompt fluence are the best predictors.}
\label{cross-validated}
\end{figure}

However, to generalize the procedure and assess the dependence of the performance on the particular choice of the training set, we use a nested 10-fold cross validation procedure repeated 100 times, see Sec Methods.
As evident there is an excellent agreement between the metrics (with a high Pearson correlation of $0.92$, bias of $2 \times 10^{-4}$ and MSE of only $5\times 10^{-3}$) of the cross-validation results (left panel of Fig. \ref{cross-validated}), and the metric obtained from the test set (right panel of Fig. \ref{ensambletraining}). This shows again that we are not overfitting.
Our results show that the accuracy of the prediction is quite stable for the majority of partitions in the training sets and test sets, although for a small number of partitions we observe higher MSE. This is indeed natural due to the large heterogeneity of the data and the relatively small sample size. It should be noted that we present results in log scale, because of the $\log$ scale transformations we applied to majority of the variables and because the best model is indeed trained in $\log (1+$z$)$. In addition, $Z=z+1$ is a more natural parametrization of the cosmological variable $z$.

In the right panel of Fig. \ref{cross-validated} we show the relative influence of the variables obtained with the ensamble of several methods (see Sec. Methods for details). The relative importance of features is an average of local importance given by a local linear approximation of predictions (see Sec. Methods for details). As we can see, the variables related to the plateau emission, flux and duration are the best predictors, thus naturally recovering the Dainotti correlation, namely the anti-correlation between the luminosity at the end of the plateau emission and the rest end time of the plateau emission (Dainotti et al. 2008, 2010, 2011a, 2013a, 2013b, 2015a, 2017a). This is expected, because prolounged emission makes easier for measuring the redshift. The third best predictor is the prompt Fluence. This implies that also the Xu \& Huang (2012) correlation, an extension of the Dainotti relation, that involves the fluence, naturally arises from the relative influence of the three most important predictors. We also point out that fluence is highly correlated with the peak flux ($r=0.65$) in the prompt emission, thus, this means that the fundamental plane relation is also independently recovered by our ML approach. This clearly shows that the application of a ML approach is able to reveal features among the variables in a multidimensional space in a completely independent way from previous approaches. Or in other words, these results prove that the Dainotti correlation plays an important role in the $z$ determination.
As shown in Stratta et al. (2018), there is a probabilist theoretical model which can explain the plateau emission and consequently the Dainotti relations: the magnetar model where an initial accretion phase not only powers the GRB jet, but also spins up a remnant magnetar, with a subsequent spin down, which gives rise to the afterglow phase. The missing exploration of the plateau emission parameters explains the much larger uncertainties and the lack of encouraging results encountered by other studies seeking ML schemes to estimate GRB redshifts (e.g. Morgan et al. 2002, Ukwatta et al. 2016), all of which have neglected the inclusion of the afterglow plateau parameters. When comparing our results to those from the random forest approach of Ukwatta et al. (2016), we note that the predictive power of our ensamble model shows a significant improvement of $61.4\%$ ($r=0.92$ vs $r=0.57$) even though we are actually using a smaller number of predictors ($10$ vs $11$) and a much smaller, but a more carefully defined GRB set (171 vs 284). We anticipate that with a larger sample we will be able to use more explanatory variables and in principle obtain more predictive results and a smaller MSE.
In our approach and Ukwatta et al. (2016) we have five common predictors ($T_{90}$, $\gamma$, $NH$, $\Gamma$, and the fluence).

\subsection*{Luminosity function and cumulative rate evolution}\label{evol}

To check whether the results we obtained can be used for deriving important astrophysical properties such as the GRB luminosity function and the cumulative rate evolution, we applied the EPL method to derive these properties from both the observed and predicted redshifts.

\begin{figure}[h!]
\includegraphics[width=0.48\linewidth,height=0.43\textwidth,angle=0,clip]{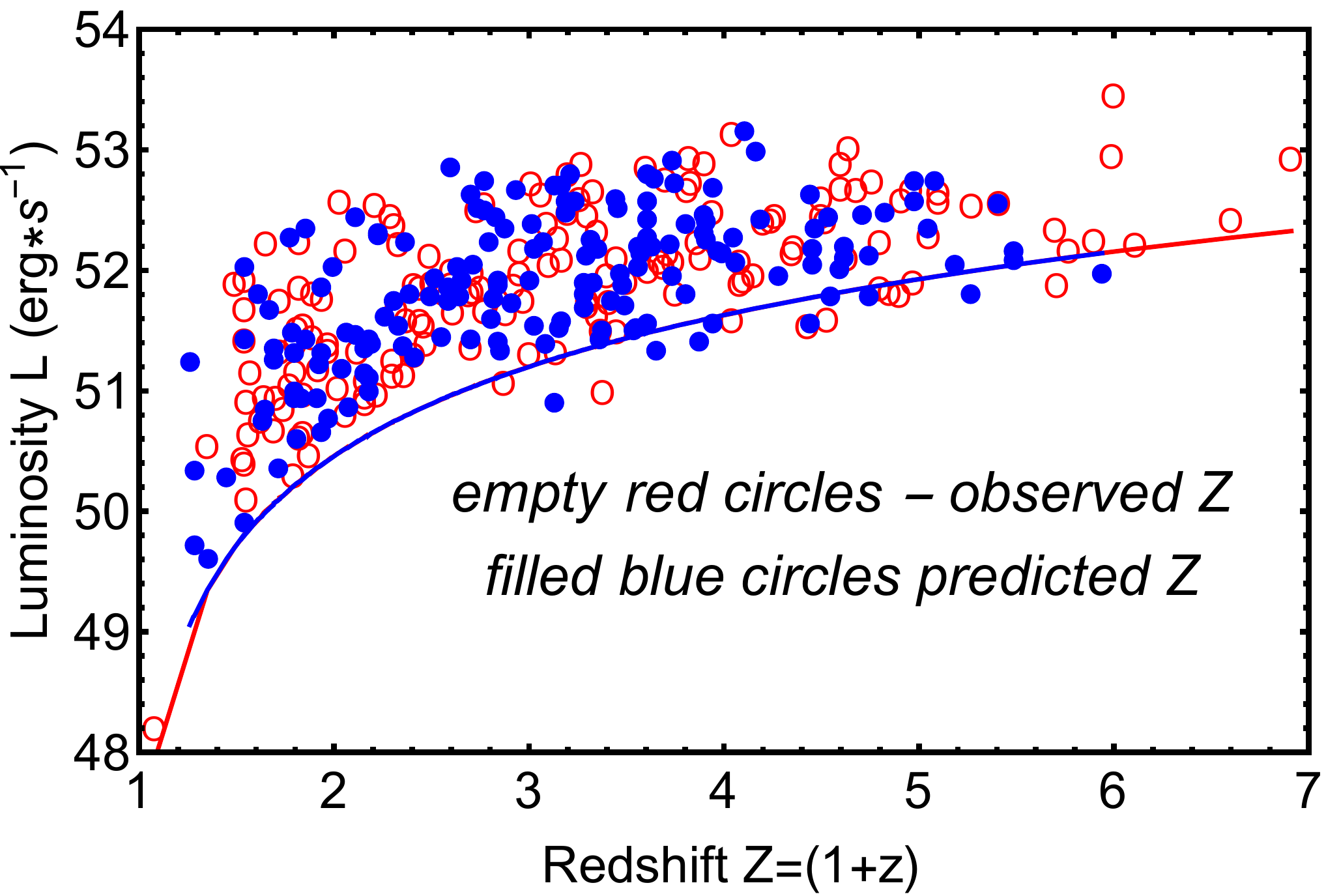}
\includegraphics[width=0.50\linewidth,height=0.45\textwidth,angle=0,clip]{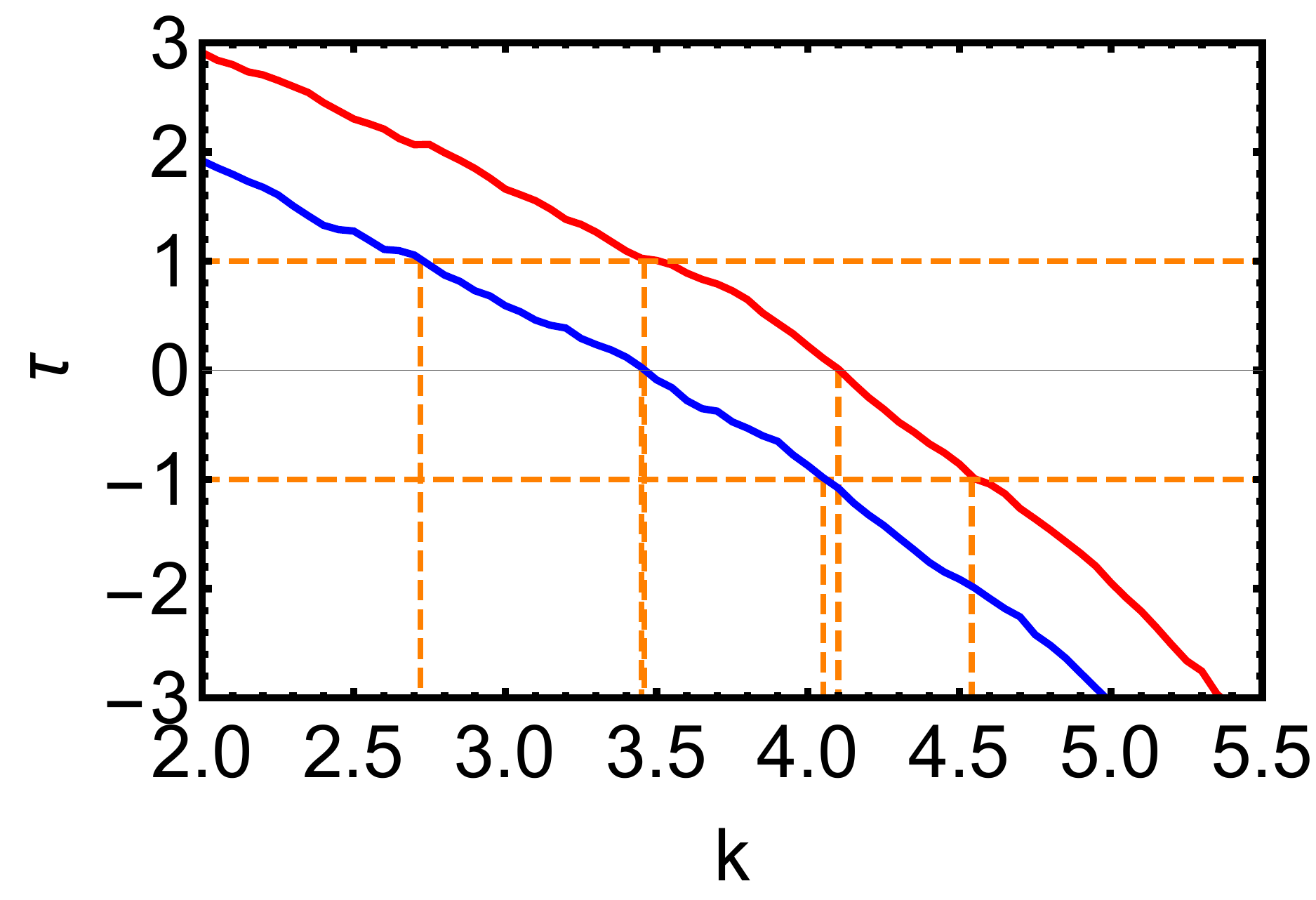}
\caption{Left panel: Luminosity of GRBs with known redshift (empty red circles) and GRBs with inferred $z$ (blue filled circles) with the same limiting luminosity using $F_{lim}=5.5 \times $10$^{-8}$ erg cm$^{-2}$. Right panel: $\tau$ versus k, where k defines the power of the $g($z$)$ function for the luminosity evolution. Red curve shows results for the GRBs with observed $z$, while the blue curves give the corresponding function for the GRBs using inferred redshift.}
\label{luminosities}
\end{figure}
	
\begin{figure}[h!]
\includegraphics[width=0.5\linewidth,height=0.4\textwidth,angle=0,clip]{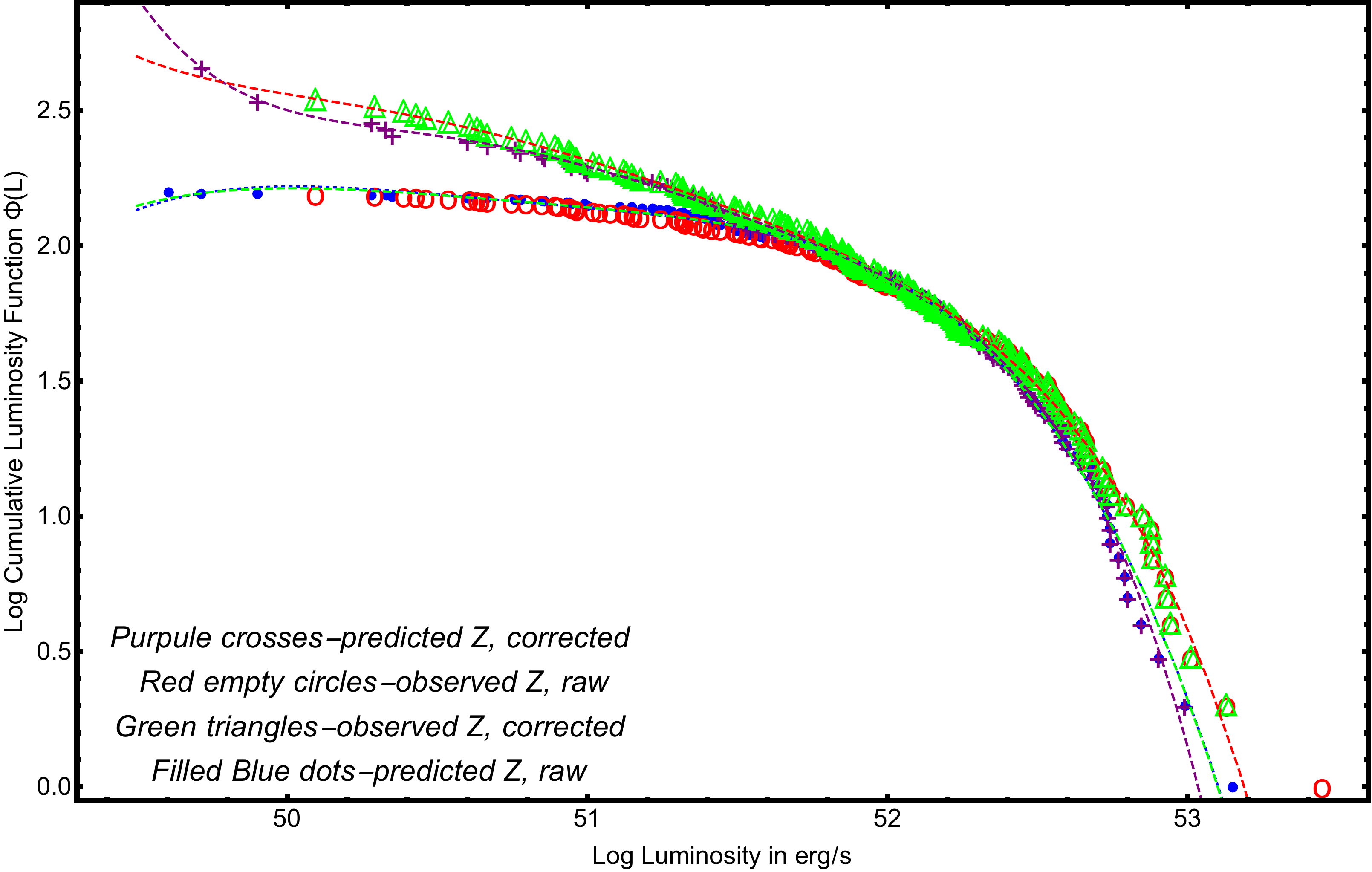}
\includegraphics[width=0.5\linewidth,height=0.4\textwidth,angle=0,clip]{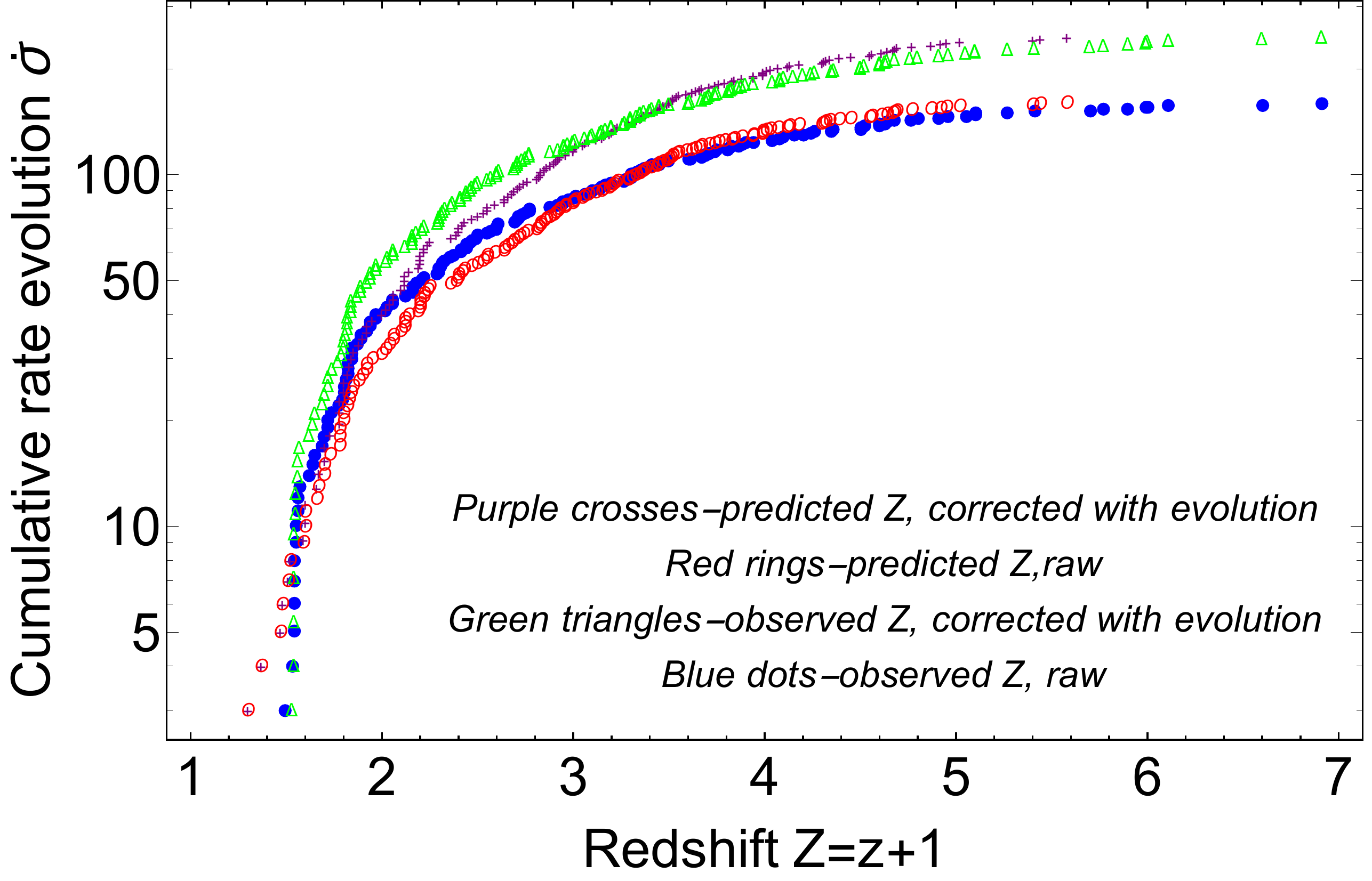}
\caption{Luminosity functions and cumulative rate evolution calculated for both observed and inferred redshifts. Note the full consistency of results for both data sets using the same flux limit.}
\label{fig:lf}
\end{figure}

The EP method requires a limiting flux, $F_{lim}$, from which we obtain the minimum observed luminosity, $L_{lim(z)}=  4 \pi D_L^2($z$) \, F_{lim} K$, where $D_L$ is the luminosity distance and $K$ is the K-correction for cosmic expansion (Bloom 2001). In the left panel of Fig. \ref{luminosities} we show the limiting luminosity; we use $K=1$ to not have fuzzy boundaries, but for an appropriate evaluation of the luminosity evolution we assign to each GRB its own K correction. We have investigated several limiting fluxes to determine a good representative value, while keeping an adequate sample size. We have finally chosen the limiting flux $F_{lim} = 5.5 \times $10$^{-8}$ \rm erg cm$^{-2}$ $s^{-1}$, for both the observed and predicted redshift, which allows $158$ GRBs.  
This limiting luminosity and luminosity of GRBs from the observed (empty red circles) and predicted redshifts (filled blue circles) are shown on left panel of Fig. \ref{luminosities}.

To estimate the luminosity function and rate evolution of our samples it is necessary to first determine whether the variable $L$ and $z$ are correlated or not (i.e. if there is a luminosity evolution). If $L$ is statistically independent of $z$, a lack of this correlation implies absence of such an evolution. The EP method uses the Kendall $\tau$ statistic (a rank measure) to test the independence of variables in truncated data and prescribes how to remove the correlation by defining new variable, $L' \equiv L/g($z$)$.

We determine $g($z$)$, which is the function that describes the evolution, so that $L^{'}$ is no longer correlated with $z$. 
The evolutionary function is commonly parametrized by a single parameter: $g($z$)=($Z$^{k} \times $Z$_c^{k})/($Z$^{k} + $Z$_c^{k})$ where $Z=$z$+1$ and a critical $Z$ is chosen as $Z_c=3.5$. We use this function for both the observed and predicted $z$.

In the right panel of Fig \ref{luminosities} we show the test statistic $\tau$ vs $k$, which have the best value within 1 $\sigma$ of $k=4.1^{+0.4}_{-0.6}$ for the GRBs with observed and $3.46^{+0.6}_{-0.7}$ for predicted redshift. Thus, both functions are consistent within a 1 $\sigma$ level.
We determined the cumulative luminosity function and rate evolution according to the EP method. Fig. \ref{fig:lf} shows these distributions for both observed (purple crosses for corrected and green rings for raw cumulative distribution) and predicted (red rings for corrected and blue dots for raw cumulative distribution) redshifts. 

We see that the resulting cumulative luminosity functions and density rate evolutions are very similar when derived from either the observed redshifts or the ML predicted redshifs. Thus, it is clear that the method used here yields reliable redshifts which will be useful towards producing complete samples as required for accurate derivations of luminosity functions.

\section*{Summary and Conclusion}\label{summary}
We have tested several ML algorithms that allowed us to determine the redshift of GRBs based on the observed characteristics of their prompt, and for the first time, the afterglow plateau emissions. We show that some of the afterglow characteristics (duration of the plateau emission and its flux) are the most powerful predictors.
A successful application of these methods to the {\it Swift} GRB catalogue can more than double the sample of GRBs with known redshift thus yielding a much larger sample which is complete to a given flux limit. This, in turn will allow us to determine the luminosity function, the luminosiy evolution and the formation rate evolution of GRBs with an unprecedented accuracy that is not possible with incomplete subsample of GRBs with known redshift. Using a sample of 171 GRBs from the {\it Swift} satellite with known redshifts we have shown that adopting ML techniques, in particular the SuperLearner package including a variety of regression methods, can provide predicted redshifts (based on only observed characteristics) that agree with the observed redshifts to a very high accuracy: a Pearson correlation coefficient of $r=0.92$, and MSE=$5\times 10^{-3}$. More importantly, the cosmological distributions (luminosity evolution, luminosity function and rate evolution) obtained from the predicted redshifts are in excellent agreement with those obtained from the observed redshifts. This result makes GRBs powerful distance estimators and is a substantial improvement compared to other methods in the literature. Indeed, we obtain an increase of $61.4\%$ in the Pearson correlation of the inferred and observed redshifts in the cross-validated samples, while previous works not including plateau emission parameters and using only one ML method achieved a correlation of $r=0.57$. 
Results from the application of this method to a large number of long GRBs will allow to solve the crescent controversy on the discrepancy between density rate evolution of long GRBs and the cosmological SFR. More importantly, application of this method to short GRBs will provide a more accurate determination of the rate of expected GWs sources associated with neutron star mergers.

\section*{Acknowledgement}
We are particular grateful to Michal Ostrowski for his precious discussion during the 
data analysis and results, to Trevor Nelson for his preliminary analysis during the summer internship program at Stanford University supported by SLAC, to Simona Scimemi for the preliminary analysis related only to prompt emission variables during the winter internship 2018 at INAF Bologna, to Monica Brora on the very preliminary analysis on random forest. We are also grateful to Macjei Bilicki and Agnieska Pollo for the useful discussion on the parameter space division. We thank particularly Douglas Figberg for his interesting discussion on selection biases related to this procedure and the way on how to solve them. M.G.D is particularly grateful for the support "JSPS Grants-in-Aid for Scientific Research KAKENHI (A) 19H00693", "Pioneering Program of RIKEN" for Evolution of Matter in the Universe (r-EMU)", "Interdisciplinary Theoretical and Mathematical Sciences Program of RIKEN" to support her stay in Riken in January-February 2019. M.G.D is also grateful to the Gilles Ferrandes and Donald Warren for their interesting discussions on the method of SuperLearner. We are very grateful for the initial discussion on the project to Dmitry Maleshev. We acknowledge useful discussion on the manuscript from Giovanni Montani and Maria Salatino.

\section*{Methods}\label{Methods}
In our analysis we have used predictive data mining, also called supervised learning. It is based on prior knowledge of a
`training' data set on which we can build models that will predict the relations in a new ``test" data set. Examples are classification and regression tools, such as k-nearest neighbors, decision trees, random forests, and gradient boosting, which allow to estimate
complicated non-linear relationships between the response and explanatory variables as well as high-order
interactions between predictors. Such non-parametric methods are very powerful when the data set contains many observations. However, they suffer from the so-called ``curse of dimensionality", which sets limits on the number of parameters one can efficiently estimate for a given sample size, which become stricter for the number of predictors which can be considered by the non-parametric methods. As a result, the fully non-parametric machine learning methods allow the use of only a limitted number of predictors, e.g. when estimating redshifts based on small GRB training sets. On the other hand, when important explanatory variables are identified by other methods, or by expert knowledge,
these methods allow the construction of efficient predictors without assumptions on the underling form of the relevant connections. In our analysis we have used supervised machine learning tools which employ regression. Within regression methods, we have used parametric, semi$-$parametric and non$-$parametric approaches. We give a brief description of each model below. 

{\bf Linear regression} for a univariate {\bf response variable}, $Y$ and a vector $X=(X_1, \ldots, X_p)$ of explanatory variables
$X$, attempts to fit the following relation: $f(x_1,\ldots,x_p)=\beta_0+\sum_{i=1}^{p}{\beta_ix_i}$.
The independent variables, namely the GRB variables of our choice, and the fitting parameters,
$\beta_i$, are determined from a training set with the Akaike (1974) information criterion (AIC).

In the {\bf semi-parametric generalized additive models (GAM), developed by Hastie, T.J., \& Tibshirani, R.J., 1986}, $Y$ is related to the predictors,
$x_1,\ldots,x_p$ through the additive model $f(x_1,\ldots,x_p)=\beta_0+\sum_{i=1}^{p} {\beta_i
f_i(x_i)}$. The functions $f$ may have some specified parametric form (e.g., a polynomial) or may be
estimated non$-$parametrically, simply as `smooth functions'. In this method, interactions between
different predictors can be included. An interaction may arise when there are two or more explanatory variables and the influence of some of them on the response variable depends on the other explanatory variables. 
In practice, interactions make it more difficult to predict the consequences of changing the value of a
variable. If any smooth functions were allowed in model fitting, then maximum likelihood estimation
of such models would result in over-fitting. Thus, the models are usually fit by penalized likelihood maximization, in which the model is modified by the addition of a penalty for each smooth function, penalizing its irregularities. 

{\bf Extreme Gradient boosting (Freedman 2001)} techniques are non$-$parametric approaches based on multiple regression
trees to increase their predictive power.
{\bf The regression tree} constructs the predictor by partitioning the data based on the values of
explanatory variables and `averaging' the value of the response variable in each element of this
partition. Partitions are formed recursively, where each of the previously created subsets of data 
is split into two parts based on the value of the selected explanatory variable. The challenging
task is determining the optimal depth of such tree (number of partition levels). 
In {\bf gradient boosting} with trees, the final predictor is constructed as a weighted sum of simple tree predictors. Compared
to the random forest methods, regression trees are not generated independently, but built on each other using residuals from the previous step, until the culmination of trees forms a stronger regression model. 

Models constructed by extreme gradient boosting are not tractable analytically. This is different for GAM, where, we obtained an analitical model composed of linear parts and two functions which describe two-way and three-way interactions between some of the predictors.

We have also tested {\bf LASSO (least absolute shrinkage and selection operator), developed by Tibshirani, R. 1996} for linear models with interaction. This method is defined as a shrinkage selection method for linear models, because it allows selection of a subset of predictors and discards the rest. Therefore, the model is more easily interpretable with a smaller number of predictors and generally has
a lower prediction error than the full model. Moreover, this model also performs
regularization, namely, introduces penalties to prevent overfitting.
Since we have experienced that each method performs optimally in different regions of the parameter space, we decided that instead of choosing a particular model, we fully exploit all methods available for use in the SuperLearner package. 

{\bf SuperLearner (Van deer Laan et al. 2007)} is an algorithm that uses internal 10-fold cross-validation to estimate the performance of multiple machine learning models, or the same model with different settings. To test the performance of our results and to estimate the prediction error we used an external layer of cross-validation, also called nested cross-validation. Namely, we performed $10$-fold cross validation repeatedly $100$ times. This type of cross-validation involves partitioning a sample of data into complementary subsets (in this case $10$), performing the analysis on 9 out of $10$ data sets (training sets) of the full data set, and validating the analysis on the remaining 1 testing set. The procedure is repeated iteratively so that each set of the $10$ will be used as test set and the remaining as training sets and the results of the prediction are averaged over the number of runs (in this case $100$). 
SuperLearner then creates an optimal weighted average of those models, e.g. an ensemble, using the test data performance. Namely, the SuperLearner provides coefficients to inform of the weight, $A_i$ or importance of each individual learner in the overall ensemble. By default the weights are always greater than or equal to $0$ and sum to $1$. 
This approach has been proven to be asymptotically as accurate as the best possible prediction algorithm tested. 

We use risk as a measure of model accuracy or performance and choose the optimal configuration by minimizing the resulting risk, which means the model is making the fewest mistakes in its prediction, i.e., it minimizes the mean-squared error in a regression model. 
We used the functions implemented in the statistical software {\it R}, particularly the SuperLearner package. We restricted the extreme gradient boosting to use at most 4-way interactions to avoid excessively complex models for the current data set. We set the number of trees to be $400$, since we have tested for the extreme gradient boosting method separately what the optimal number of trees is. Increasing the number of trees beyond this point results in no further increase in the Pearson correlation coefficient which indeed remains constant. 
To estimate the contribution of each predictor we use the relative importance of features which is an average of local importance given by a local linear approximation of prediction. Specifically, for every observation, we create a corresponding synthetic data by adding Gaussian noise. Next, we construct an approximate change in prediction through a linear model prediction, P= X$B$ were $B$ are fit coefficients, and we define the local relative importance of feature $i$ over all sample points by $R_i =|B_i|/\sum|B_i|$. 

\section*{References}
\begin{quote}
\begin{verbatim}

Abbott, B. P. et al. 2017, Nature, 551, 85
Akaike, H. 1974, IEEE Trans. Aut. Contr., 19, 716 
Amati, L., et al. 2002, A & A, 390, 81
Atteia, J. L., et al. 2003, A\&A, 407, L1
Collazzi & Schaefer 2008, ApJ, 688, 1, 456.
Dainotti, M. G., et al. 2008, MNRAS, 391L, 79
Dainotti, M. G., et al. 2010, ApJL, 722, L215
Dainotti, M. G., et al. 2011a, ApJ, 730, 135
Dainotti, M. G., et al. 2013a, ApJ, 774, 157
Dainotti, M. G., et al. 2013b, MNRAS, 436, 82
Dainotti, M. G., et al. 2015a, ApJ, 774, 157 
Dainotti, M. G., et al. 2015b, MNRAS, 451, 4
Dainotti, M. G., et al. 2016, ApJL, 825, 6
Dainotti, M. G., et al. 2017a, A&A, 600, 11
Dainotti, M. G., & Del Vecchio, R., 2017b, NAREV, 77C, 23
Dainotti, M. G., et al. 2017c, ApJ, 848, 88
Dainotti, M. G., 2018a, A & A, art. id 4969503, 31 
Dainotti, M. G., et al. 2018b, PASP, 30, 987.
Efron, B. & Petrosian, V., 1992, ApJ, 399, 345
Efron, B. & Petrosian, V. 1999, J. Am. St. Assoc., 94, 824 
Friedman, J. H. 2001, The Annals of Statistics, 29, 5, 1189.
Gehrels, N., et al., 2004, ApJ, 611, 1005
Ghirlanda et al. 2004, ApJ, 616, 331 
Ghirlanda, G., et al. 2016, A & A, 594, id.A84
Guetta & Piran 2005, A&A, 435, 2, 421
Guetta & Piran 2006, A&A, 453, 3, 2006, 823
Kocevski, D., & Liang, E.  2006, ApJ, 642, 371
Kouveliotou, C., et al., 1993, ApJ, 413, 2, L101
Jakobson et al. 2006, A&A, 447, 3, 897
Hastie, T.J., Tibshirani, R.J., 1986, Stat. Sci. 1, 297.
Lynden-Bell, MNRAS, 155, 119, 1971.
Lloyd, N. M., & Petrosian, V. 1999, ApJ, 511, 550 
Lloyd, N. M. et al 2000, ApJ, 534, 227
Lloyd, N. M. et al. 2002, ApJ,  574, 554
Madau, P. & Dickinson, M. 2014, ARAA, 52, 415
Mazets & Golenetskii 1988
Maloney & Petrosian 1999, ApJ, 518, 1, 32.
Metzger & Berger 2012, ApJ, 746, 1, 48, 15
Morgan, et al. 2012, ApJ, 746, 170 
Nakar, E., & Piran, T. 2005, MNRAS, 360, 1 
Pescalli, A., et al. 2016, A & A, 587, A40
Petrosian, V. 1992, New York:Springer, 173
Petrosian V., et al. 2015, ApJ, 806, 44
Petrillo e al. 2013, ApJ, 767, 2, 140, 6.
Porciani, C., & Madau, P. 2001, ApJ, 548, 522
Salvaterra R. 2009, Nature, 461, 7268, 1258.
Singal et al. 2011, ApJ, 743, 2, 104, 13.
Stratta, G., et al. 2018, ApJ, 869, 155
Schmidt, M. 1968, ApJ, 151, 393
Troja, E. et al. 2017, Nature, 55, 71
Tsvetkova, A., et al. 2017, AJ, 850, 2, art. Id. 161, 27
Tibshirani, R. 1996, JRSS, Series B, Wiley. 58, 1, 267.
Ukwatta, T. N., et al. 2016, MNRAS, 458, 3821
van der Laan, M. et. al. 2007, Working Paper 222.
Xu, M. & Y. F. Huang, A & A 2012, 538, 134
Yonetoku, D., et al. 2004, ApJ, 609, 2; 609, 935
Yonetoku, D., et al. 2014, ApJ, 789, 65
Yu, H., et al. 2015, ApJS, 218, 13
Zhang, G.Q., & Wang, F.Y. 2018, ApJ, 852, 1

\end{verbatim}
\end{quote}

\end{document}